\documentclass[aps,prb,twocolumn,groupedaddress,showpacs]{revtex4}

\usepackage[bookmarks]{hyperref}
\usepackage[dvips]{graphicx}
\usepackage{amsmath}
\usepackage{amstext}
\usepackage{exscale}
\usepackage{changebar}
\usepackage[T1]{fontenc}
\usepackage{bm}
\usepackage{bbm}

\usepackage{version}

\usepackage{latexsym}

\usepackage[tight]{subfigure}

\usepackage{amssymb}
\usepackage{amsbsy}

\renewcommand{\epsilon}{\varepsilon}

\newcommand{\gfbraket}[1]{\langle\!\langle #1  \rangle\!\rangle}

\newcommand{\elcre}[2]{ c^{\dagger}_{#1,#2}}
\newcommand{\elann}[2]{ c_{#1,#2}}

\newcommand{\vk}{{\bm k}}

\newcommand{\Imag}{\mathrm{Im}}
\newcommand{\Real}{\mathrm{Re}}

\newcommand{\hc}{\mathrm{h.c.}}

\includeversion{commentst1} 

\begin{document}

\title{Low-energy kink in the nodal dispersion of
  copper-oxide superconductors: Insights from Dynamical Mean Field Theory} 
\author{Johannes Bauer${}^1$ and Giorgio Sangiovanni${}^2$}
\affiliation{${}^1$Max-Planck Institute for Solid State Research, Heisenbergstr.1,
  70569 Stuttgart, Germany} 
\affiliation{${}^2$ Institute of Solid State Physics, Vienna University of Technology, 1040 Vienna, Austria}
\date{\today} 
\begin{abstract}
Motivated by the observation in copper-oxide high-temperature superconductors,
we investigate the appearance of kinks in the electronic dispersion due to
coupling to phonons for a system with strong electronic repulsion.  
We study a Hubbard model supplemented by an electron-phonon coupling of
Holstein type within Dynamical Mean Field Theory (DMFT) utilizing Numerical
Renormalization Group as impurity solver.  Paramagnetic DMFT solutions in the
presence of large repulsion show a kink only for large values of the
electron-phonon coupling $\lambda$  or large doping and, contrary to the
conventional electron-phonon theory, the position of such a kink can be shifted to
energies larger than the renormalized phonon frequency $\omega_0^r$.
When including antiferromagnetic correlations we find a
stronger effect of the electron-phonon interaction on the electronic
dispersion due to a cooperative effect and a visible kink at $\omega_0^r$,
even for smaller $\lambda$. Our results provide a scenario of a kink position
increasing with doping, which can be related to recent photoemission
experiments on Bi-based cuprates. 
\end{abstract}
\pacs{74.72.-h,71.38.Mx,71.10.Fd,71.20.Tx,71.10.-w,63.20.Kr}

\maketitle

\section{Introduction}

One of the manifestations of a strong coupling between condensed matter
electrons and lattice vibrations is the occurrence of kinks (abrupt changes of
the slope) in the electronic dispersion relation. Such kinks occur at a certain energy
associated with a phonon mode and their strength is directly related to the
electron-phonon coupling constant $\lambda$. Therefore the observation of a
kink, for instance, in angle resolved photoemission (ARPES) data, is often taken
as an indication that a system possesses a strong electron-phonon coupling and
its properties are likely to be strongly influenced by this. However, also
coupling to other bosonic modes possibly of purely electronic origin, can have
the same effect on the dispersion relation. Moreover, in addition to an
electron-phonon coupling different interactions important at low energy can
compete or cooperate with it, which may modify the picture. Therefore, the
experimental observation must be analyzed carefully before conclusions about
the origin of a kink can be made.

Prominent examples for the observation of kinks come from the field of cuprate
superconductors. Studies by Lanzara {\it et al.} \cite{Lea01} collected
evidence for this in different compounds.  
One interpretation of these experiments is that the dispersion kink along the
nodal direction of the Brillouin zone is associated to a longitudinal in-plane
bond-stretching mode in which two of the four oxygen atoms surrounding the
copper move inwards (``half-breathing'' mode) \cite{dastuto2}. The same phonon
mode indeed displays a sizable renormalization in inelastic neutron scattering
measurements \cite{pint2}, pointing towards a strong coupling to carriers. 
This conclusion is further supported by Iwasawa {\it et al.} \cite{Iea08}, who
carefully studied the effect of an ${}^{16}$O-${}^{18}$O isotope substitution
and concluded that their observation is compatible with the ``half-breathing''
mode.  
On the other hand, recent work by Dahm {\it et al.} \cite{dahmNatPhys}
emphasized the connection between the nodal kink and spin-fluctuations, an
alternative scenario.\cite{Jea01,MEB01,CN04,Esc06}
Hence, presently no general consensus exists on this important issue of
high-temperature superconductors. \cite{communication} 
A review article covering numerous aspects of this topic has been recently
published by A.~S.~Mishchenko \cite{mishchenko_rev2}. 

One of the aspects of the nodal kink which is particularly puzzling is the
doping dependence of its position. Early studies indicated that the nodal kink
occurs at a more or less constant energy position \cite{kink2}. A more recent
study by Kordyuk {\it et al.} \cite{kordyukPRL97} showed however that in LSCO
the kink position $\omega_k$ first increases with doping, exhibits a maximum around
optimal doping, and eventually decreases again, in a dome-like shape.  
In Bi-2212 the situation is even more intricate as the non-monotonic
dependence is replaced, at low temperatures, by a peculiar increase upon
increasing doping. 
Such a pronounced doping and temperature dependence is a very strong
indication that the nodal kink in cuprates cannot be described within the
textbook picture of conventional metals with electron-phonon interaction. 

The standard description of effects of electron-phonon coupling is 
based on perturbation theory in the electron-phonon coupling constant assuming
that electrons can be described as a weakly interacting Fermi liquid
\cite{eng-sch}. 
Then, to lowest order in the coupling the magnitude of a kink, $m_k$, in the
electronic dispersion relates to the dimensionless electron-phonon coupling
$\lambda$ as $m_k=1+\lambda$.
A clear kink, however, points to a relatively strong electron-phonon coupling, for
which one expects effects beyond the lowest order diagrams and possibly vertex
corrections to become very important.  Vertex 
corrections are beyond the Migdal-Eliashberg formalism, i.e. beyond the most
common diagrammatic theory for the electron-phonon
interaction. \cite{BZ98,MHB02,massimo4}.  
If on top of that one takes also the electron-electron interaction and Mott
physics into account, as in any minimal model for high-temperature cuprate
superconductors, a quantitative description of the nodal kink within
conventional theories seems inconceivable. 

Substantial progress in understanding strongly coupled electron-phonon systems
has been developed in recent years based on dynamical mean field (DMFT)
methods \cite{GKKR96} and cluster extensions \cite{review05}.  
The advantage of these approaches is that electron-electron and electron-phonon
interactions are treated non-perturbatively and on equal footing
\cite{KMH04,StJ2,KHE05,StJ3,macridin,werner-ph,Bau10,BH10}. 
One major insight of these studies is that the electron-phonon interaction in
strongly correlated metals cannot be understood with conventional tools.  
In particular, the polaronic signatures in photoemission spectra are more
complicated than what suggested by perturbation theory on top of a Fermi
liquid solution \cite{eng-sch}.  
This is also in line with the conclusions of other studies which do not rely
on the dynamical mean field approximation \cite{mishchenko,olle2,threeband}. 

Even though DMFT misses a crucial part of the non-local physics of the
two-dimensional copper-oxygen planes, it is interesting to see whether some
aspects of the nodal kink puzzle posed by recent experiments can be understood
already within the purely local picture of the interplay between the
electron-electron repulsion and the retarded phonon-mediated attraction given
by DMFT. 
This is the motivation for the present study, which focuses on a description of
the kinks in the electronic dispersion due to phonons in the presence of
strong electronic correlations. 
We employ the numerical renormalization group (NRG) \cite{Wil75,BCP08} as DMFT
impurity solver.  NRG is known to have a good low energy resolution at low temperatures
\cite{PPA06,WD07,BCP08}, and one can calculate spectral functions directly on
the real axis, which makes it the method of choice for the analysis of
low-energy kinks.   

We first analyze paramagnetic (PM) solutions in the vicinity of half-filling and
zero temperature.  In order to see the effect of antiferromagnetic (AF)
correlations, which are expected to be important in the cuprates, we also
perform calculations in the commensurate AF state. 
The latter have been shown to successfully mimic the effect of
antiferromagnetic correlations and exchange coupling
$J$. \cite{ZPB02,BH07c,AFDMFT} 
Away from half filling the Hubbard model can show phase separation or
incommensurate order, hence, the analysis of the results must be done very
carefully.

Our study is based on a model Hamiltonian, Hubbard Holstein model, of the form
\begin{eqnarray}
  \label{hubholham}
  H&=&-\sum_{i,j,{\sigma}}(t_{ij}\elcre i{\sigma}\elann
j{\sigma}+\hc)+U\sum_i\hat n_{i,\uparrow}\hat n_{i,\downarrow} \\
&&+\omega_0\sum_ib_i^{\dagger}b_i+g\sum_i(b_i+b_i^{\dagger})\Big(\sum_{\sigma}\hat
n_{i,\sigma}-1\Big).
\nonumber
\end{eqnarray}
$\elcre i{\sigma}$ creates an electron at lattice site $i$ with spin $\sigma$,
and $b_i^{\dagger}$ a phonon with oscillator frequency $\omega_0$,
$\hat n_{i,\sigma}=\elcre i{\sigma}\elann 
i{\sigma}$. We include nearest neighbor hopping $t_{ij}$ with magnitude
$t$. The electrons interact locally with strength $U$, and 
their density is coupled to a local phonon mode with coupling constant
$g$. We have set the ionic mass to $M=1$ in Eq. (\ref{hubholham}). 
We define $\lambda=\rho_0 2g^2/\omega_0$ in terms of bare parameters where
$\rho_0$ is the electronic density of states at the Fermi energy.

For the non-interacting density of states $\rho_0(\epsilon)$  we take
the semi-elliptic form $\rho_0(\epsilon)=2\sqrt{D^2-\epsilon^2}/(\pi D)^2$. 
In all the results we present here we take the value $W=2D=4t=4$ for the
bandwidth. The DMFT and NRG calculations are carried out as
described in detail in Refs. \onlinecite{ZPB02,KMH04,BH07c,Bau10,BH10}.

The Paper is organized as follows: In section \ref{sec:kinkPM} we discuss the
occurrence of kinks in the electronic dispersion in the normal, paramagnetic
state. We first look at results for $U=0$ and then analyze the situation for
large $U$. In section \ref{sec:kinkAFM} we show results when the effect of antiferromagnetic
correlations are included in the calculations. In Section \ref{sec:Concl} we conclude and
relate our results to the experimental observations.

\section{Kinks in the normal state}
\label{sec:kinkPM}
Before analyzing the strong coupling situation we briefly recall the description of
electron-phonon kinks in the weak coupling theory (for a review see, e.g.,
Refs. \onlinecite{CLZSDN05,ollerev_eph}). 
In the paramagnetic state the Green's function reads,
\begin{equation}
  G_{\vk}(\omega)=\frac{1}{\omega-\epsilon_{\vk}+\mu-\Sigma(\omega)},
\label{gfct}
\end{equation}
where we assume a $\vk$-independent self-energy.\cite{MV89} Generally, the
self-energy has contributions due to electron-phonon 
coupling $g$ and electron-electron repulsion.
The electron spectral function is given by $\rho_{\vk}(\omega)=-\Imag
G_{\vk}(\omega)/\pi$ and the full (interacting) dispersion relation
$E_{\vk}=E(\epsilon_{\vk})$ by the maxima of $\rho_{\vk}(\omega)$. This 
corresponds usually to the poles of Eq. (\ref{gfct}), i.e. the solution of
\begin{equation}
  E_{\vk}-\epsilon_{\vk}+\mu-\Real\Sigma(E_{\vk})=0.
\label{dispeq}
\end{equation}
We define a kink as (more or less) abrupt change in the slope of
$E(\epsilon_{\vk})$,  which consequently must be found in $\Real\Sigma(\omega)$.

\begin{figure}[!t]
\centering
\includegraphics[width=0.35\textwidth]{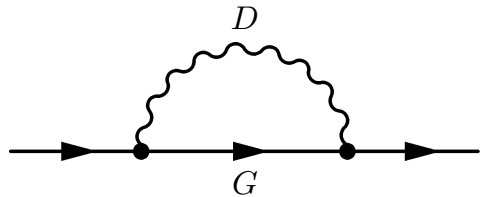}
\caption{Lowest order diagram for the electronic self-energy due to electron
  phonon coupling.}        
\label{g2diagram}
\end{figure}
\noindent
In a calculation to order $g^2$ (depicted in Fig. \ref{g2diagram}) one finds
that the $\Real\Sigma(\omega)$ is roughly linear for small $|\omega| \ll \omega_0$,
$\Real\Sigma^g(\omega)=-\lambda \omega$, strongly peaked at
$\omega\simeq \omega_0$ (logarithmic divergence), and small for $|\omega| \gg \omega_0$. This yields for the low energy
dispersion $E^l_{\vk}=\epsilon_{\vk}/(1+\lambda)$ for $|E_{\vk}|\ll\omega_0$
and  $E^h_{\vk}=\epsilon_{\vk}$ for $|E_{\vk}|\gg\omega_0$. The magnitude of the kink is 
characterized by the ratio of the high energy over low energy slope, which
gives $m_k=1+\lambda$ as mentioned 
earlier. This quantity is equal to the inverse of the quasiparticle
renormalization factor $Z$ and for a $\vk$-independent self-energy to the
effective mass $m^*/m_0=Z^{-1}$.  
The free phonon propagator is sharply peaked at $|\omega|=\omega_0$. Due to the
coupling to the electrons the peak can be shifted to a renormalized value
$\omega_0^r$ and broadened. For the phonon properties, we consider the function 
$B(\omega)=\gfbraket{b;b^{\dagger}}_{\omega}$, which can be calculated in the
NRG from the matrix elements and excitations. The corresponding spectral function is
$\rho_b(\omega)=-\Imag B(\omega)/\pi$.

\subsection{Results at $U=0$}
\begin{figure}[!thbp]
\centering
\subfigure[]{\includegraphics[width=0.23\textwidth]{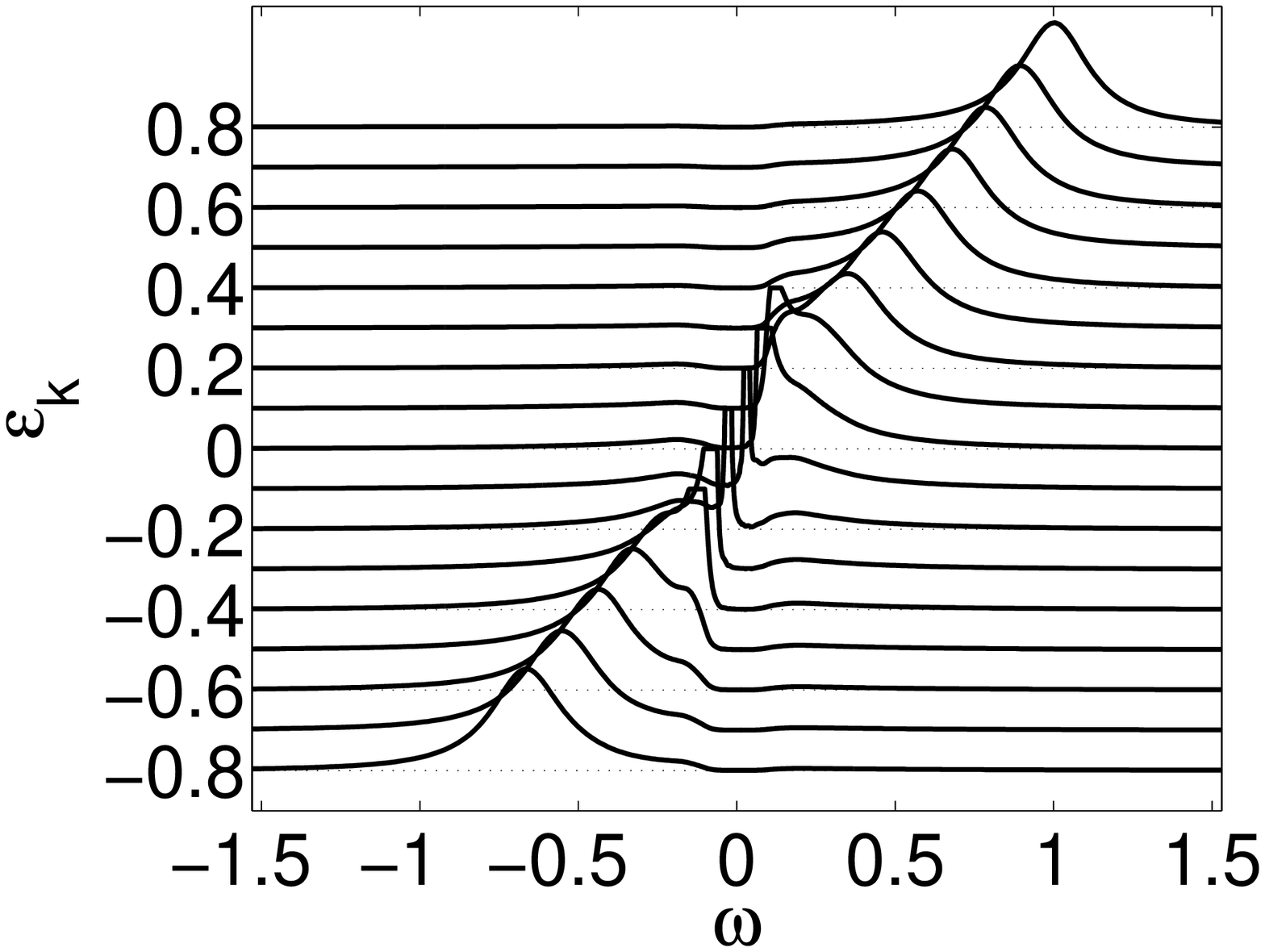}}  
\subfigure[]{\includegraphics[width=0.23\textwidth]{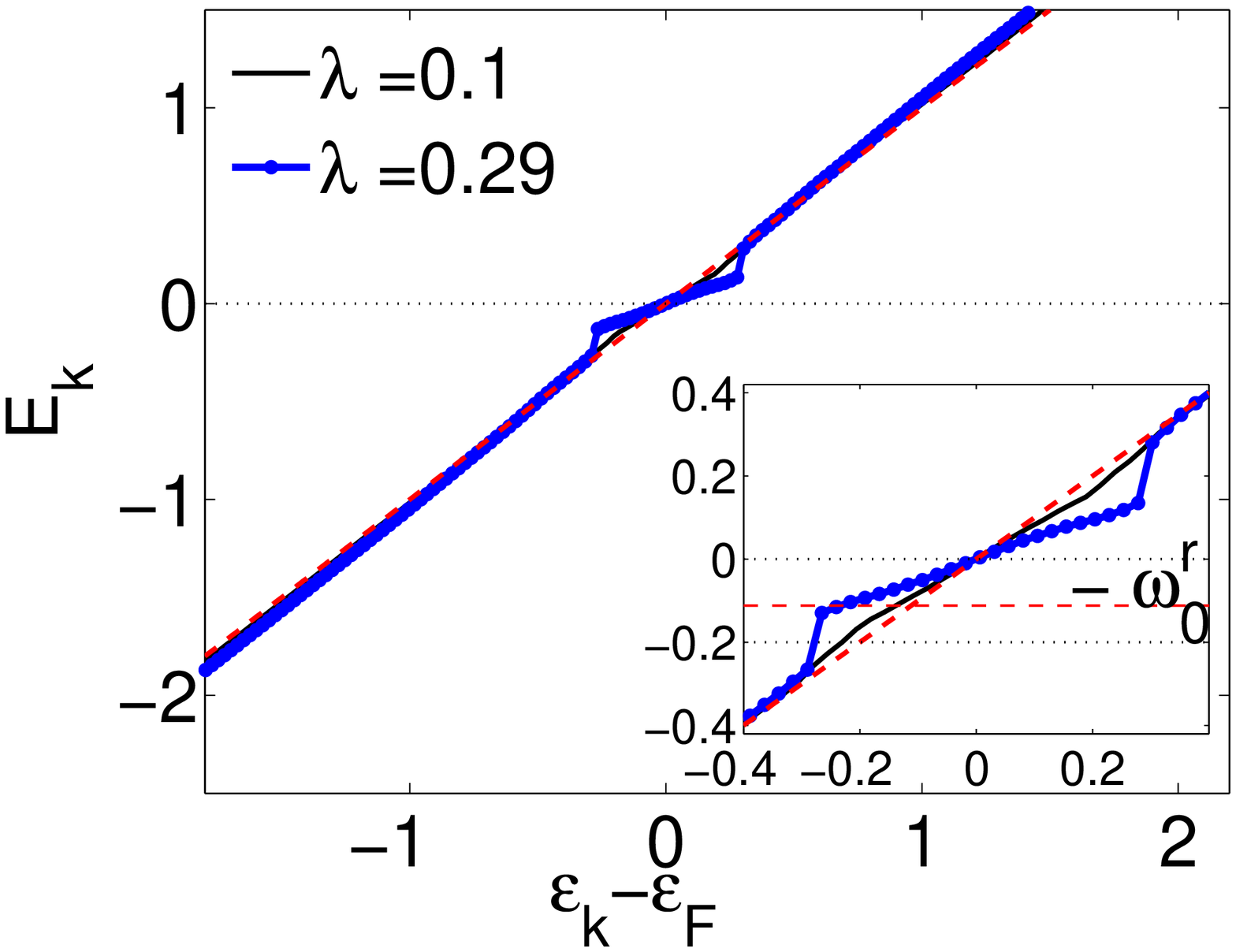}}
\subfigure[]{\includegraphics[width=0.23\textwidth]{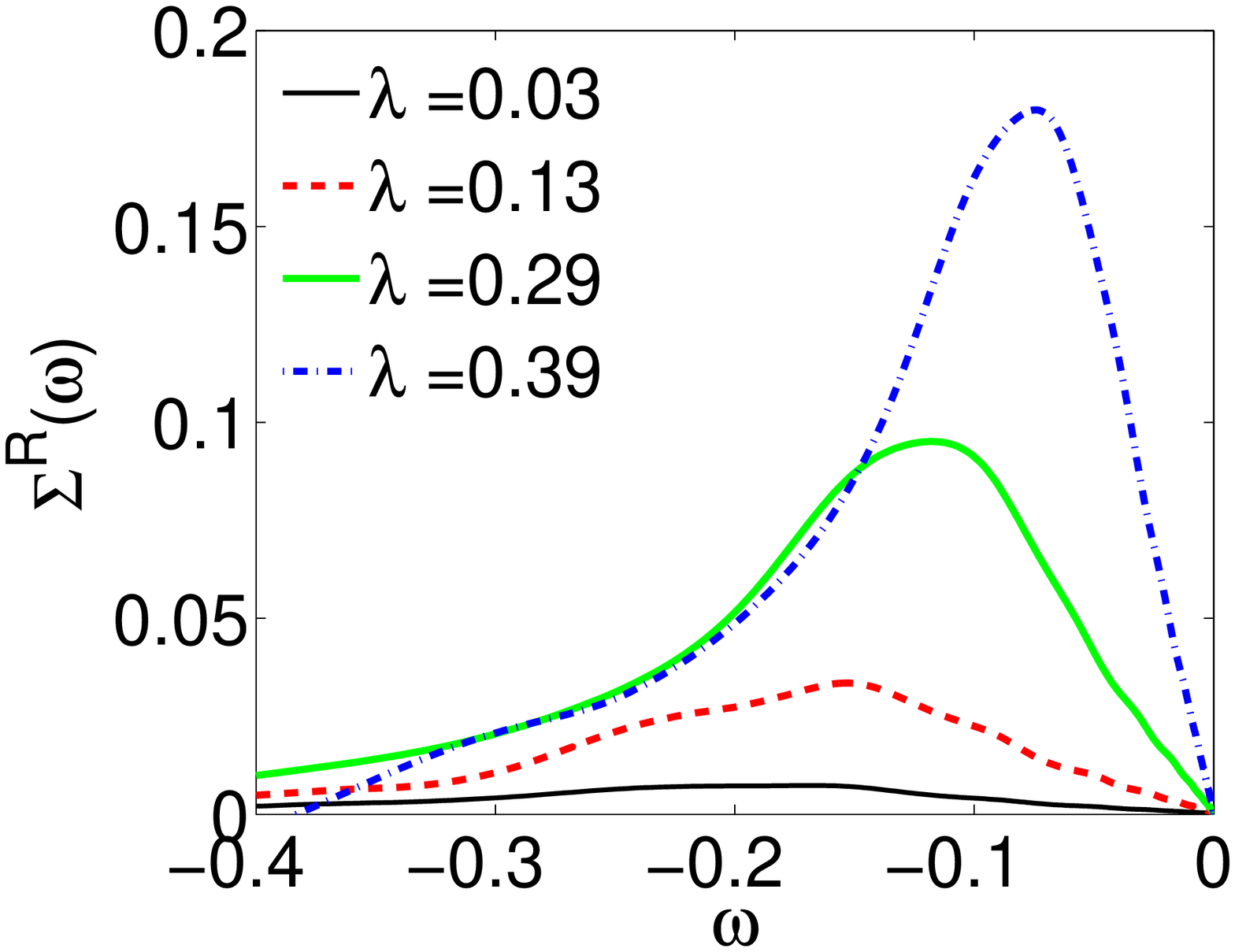}}
\subfigure[]{\includegraphics[width=0.23\textwidth]{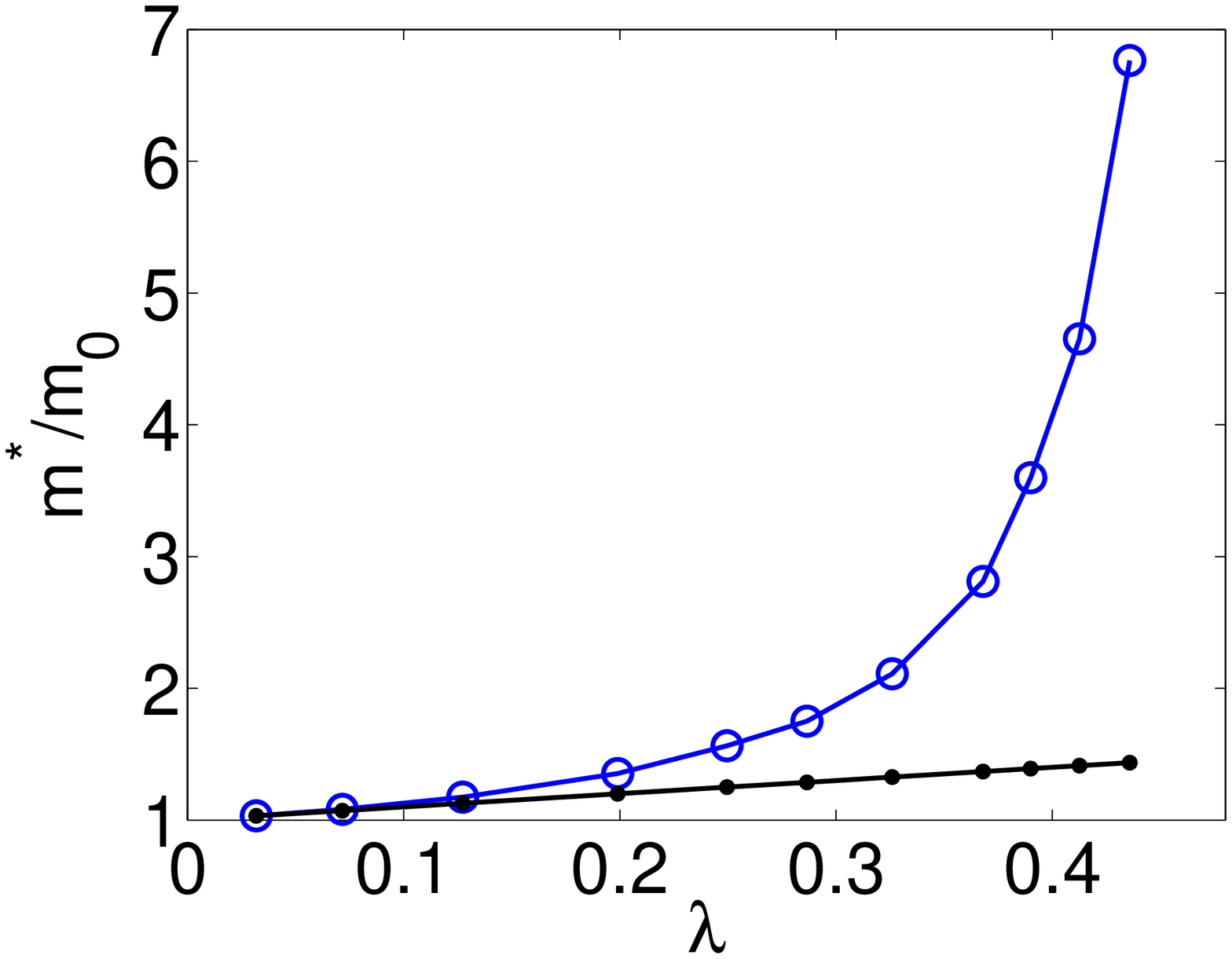}}
\caption{(Color online) PM calculation: (a) The $\epsilon_{\vk}$ resolved spectral function for $U=0$ and
  $\lambda=0.29$. (b) The dispersion relation $E_{\vk}$ as a function of
  $\epsilon_{\vk}$  for $U=0$ and different values of electron-phonon
  coupling. The dashed line  is for free electrons,
  $E_{\vk}=\epsilon_{\vk}$. (c) The real part of the self-energy $\Sigma^{\rm 
    R}(\omega)$ (with $\Sigma^{\rm  R}(0)$ subtracted) as a function of $\omega$  for $U=0$ and various values of
  $\lambda$. (d) The DMFT result for the effective mass (circles) compared
  with the second order perturbation theory result $1$$+$$\lambda$ (straight line).}        
\label{resultsU0}
\end{figure}

We first want to compare our DMFT-NRG results with the simple minded theory
and consider the case $U=0$. 
We use $\omega_0=0.2 t$, and a filling factor of $n=0.9$.
In Fig. \ref{resultsU0} (a) an example of the $\epsilon_{\vk}$ resolved
spectral function $\rho_{\vk}(\omega)$ is plotted for $\lambda=0.29$.

\noindent
We can see sharp quasiparticle peaks at low energy and broad peaks from
certain energy on, $\omega_k\simeq\omega_0^r$, where the kinks in the
dispersion occurs (see Fig. \ref{resultsU0} (b)).  
At this energy the imaginary part of the electronic self-energy sets in, and
consequently the electronic spectrum becomes much broader. 

The dispersion relation $E_{\vk}$ is obtained numerically from the peak of
position of $\rho_{\vk}(\omega)$ as a function of
$\epsilon_{\vk}$.  It is shown for two values of $\lambda$ in
Fig. \ref{resultsU0} (b). 
We find the picture of a kink as inferred from the lowest order $g^2$ diagram,
i.e., the low energy dispersion is renormalized and at the (renormalized)
phonon scale $\omega_0^r$ it reverts back to the non-interaction slope. The magnitude of
the kink $m_k$ can be inferred from the low-energy slope.
For $\lambda=0.29$,  we find the value $m_k=1.75$, whilst the corresponding
value from the second order theory is with $m_k=1.29$ substantially smaller.
Comparing with part (a), we can also see how the discontinuity in the  
dispersion relation comes about through the change in maximum position. 

\begin{figure}[!thbp]
\centering
\includegraphics[width=0.45\textwidth]{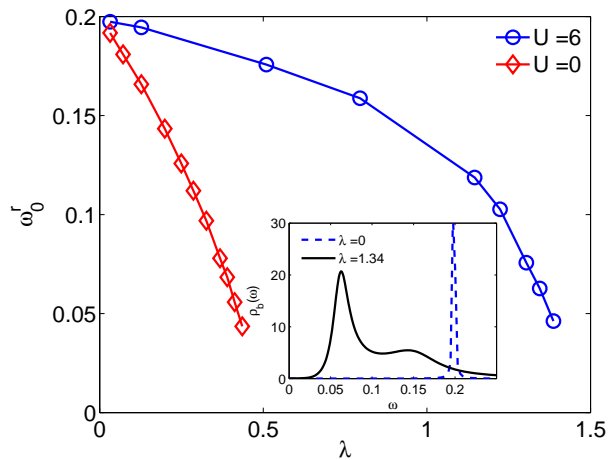}
\caption{(Color online) PM calculation: The renormalized phonon frequency $\omega_0^r$ as a function of $\lambda$
  for $U=0$ and $U=6$. The inset shows the phonon spectral function
  $\rho_b(\omega)$ for $U=6$ and $\lambda=0$ and $\lambda=1.34$.}       
\label{phonsoftU0}
\end{figure}

The corresponding behavior of the real part of the self-energy, $\Sigma^{\rm
  R}(\omega)$, is shown in Fig. \ref{resultsU0} (c).
We can see how the low energy slope increases with $\lambda$ together with a
peak at the renormalized phonon frequency $\omega_0^r$, smaller
than the bare value $\omega_0=0.2$. The renormalized phonon frequency
$\omega_0^r$ as a function of $\lambda$  is shown explicitly in
Fig. \ref{phonsoftU0}. 

\noindent
A comparison for $1+\lambda$ with the DMFT results for a series of values can
be found in Fig. \ref{resultsU0} (d). 
There are quantitative deviation already at relatively small values of
$\lambda$ and very notable ones for larger $\lambda$.\cite{massimo4}
Hence, for negligible $U$ a very large kink is obtained already for values of
the order $\lambda\sim 0.4$.
For larger values of $\lambda$ a bipolaronic instability occurs where local
pairs are formed. At half filling above a critical coupling stable solutions
of a bipolaronic insulator can be found as in previous DMFT
calculations. \cite{MHB02,massimo4,KMH04,hhdoping} 
However, in the doped system the calculations become unstable, 
and results oscillate between solutions with small and large double
occupancy. Paramagnetic DMFT calculations can not provide a suitable ground
state solution in these situations.

\subsection{Results for finite $U$} \label{sec:PM}
We now turn to the situation with finite Coulomb repulsion, where we have
contributions to $\Sigma(\omega)$ from both interaction terms.
Therefore, already for $\lambda=0$ electrons are renormalized due to scattering processes, which leads to an 
effective mass enhancement and finite lifetime. 
For small values of the Coulomb interaction, $U\le D$, qualitatively similar 
results to the ones discussed for $U=0$ can be found for the dispersion
relation and the occurrence of kinks.  

For larger values of $U$ the situation changes. One important thing to
consider is that we have Holstein phonons, i.e. lattice vibrations which
couple to \emph{local} fluctuations of the electronic density. 
Such phonon modes are in direct competition with the Hubbard repulsion, which
does exactly the opposite: it tries to freeze density fluctuations out. 
As a consequence, larger values of $\lambda$ (compared to $U=0$) are necessary
for observing effects on the dispersion.  
Furthermore, for values of $U$ relevant for strongly correlated materials,
$U\ge W$, the high energy slope does not revert to the non-interacting one for 
$\epsilon_{\vk}\in(-D,D)$. 
Thus, when characterizing the kink it does not make sense to compare with the non-interacting slope. 
Instead one can compare the high and low energy slope within one calculation. 

\begin{figure}[!thbp]
\centering
\includegraphics[width=0.45\textwidth]{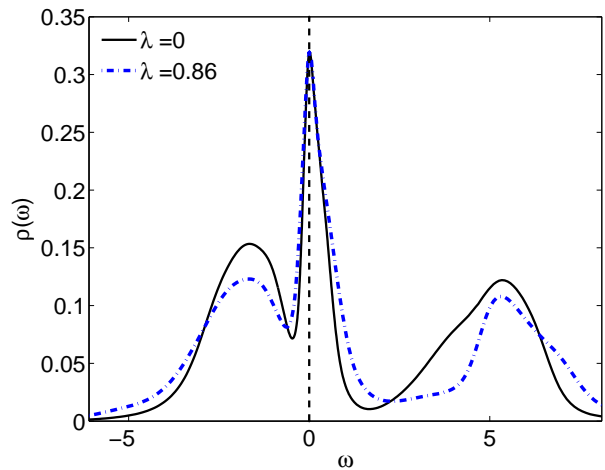}
\caption{(Color online) PM calculation: The local electronic spectral function for $U=6$ and $n=0.9$.}       
\label{specU6}
\end{figure}

In order to be in a parameter regime roughly suitable for cuprate physics, we choose
$n$=$0.9$ (10$\%$ hole doping) and $U$=$6 \! > \! 
U_{c2}$, $U_{c2}\simeq 5.88$ being the critical value for the metal-insulator transition
at half-filling ($n$=$1$).\cite{Bul99}  
We first report the  local electronic spectral function for $\lambda=0$ and
$\lambda=0.86$ in Fig. \ref{specU6}.
The well-known three-peak structure can clearly be distinguished, with lower
and upper Hubbard bands and a quasiparticle band at the Fermi level
$\omega\!\!=\!\!0$. The spectrum is not symmetric because of the finite hole
doping. 

\begin{figure}[!thbp]
\centering
\subfigure[]{\includegraphics[width=0.23\textwidth]{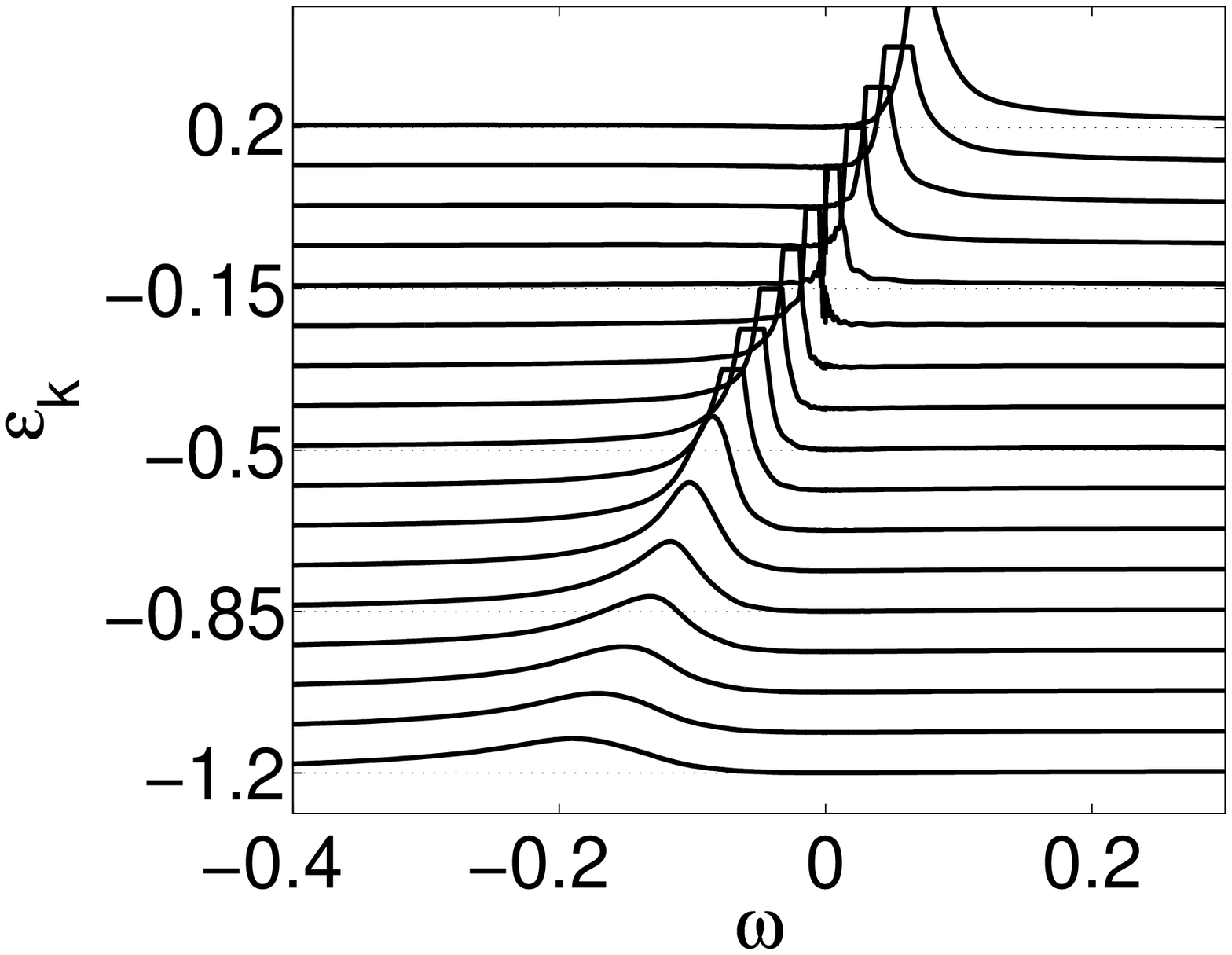}}
\subfigure[]{\includegraphics[width=0.23\textwidth]{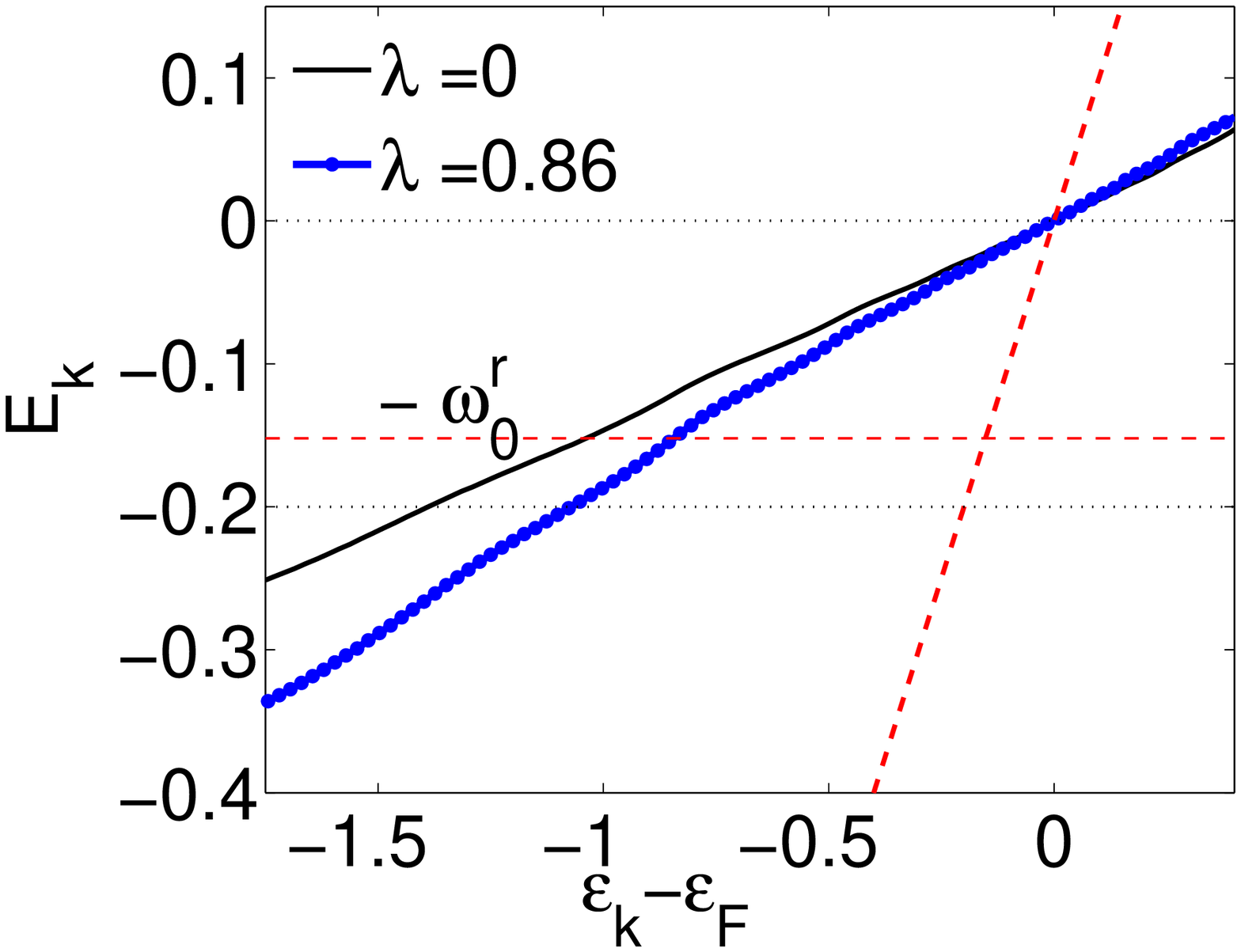}}
\subfigure[]{\includegraphics[width=0.23\textwidth]{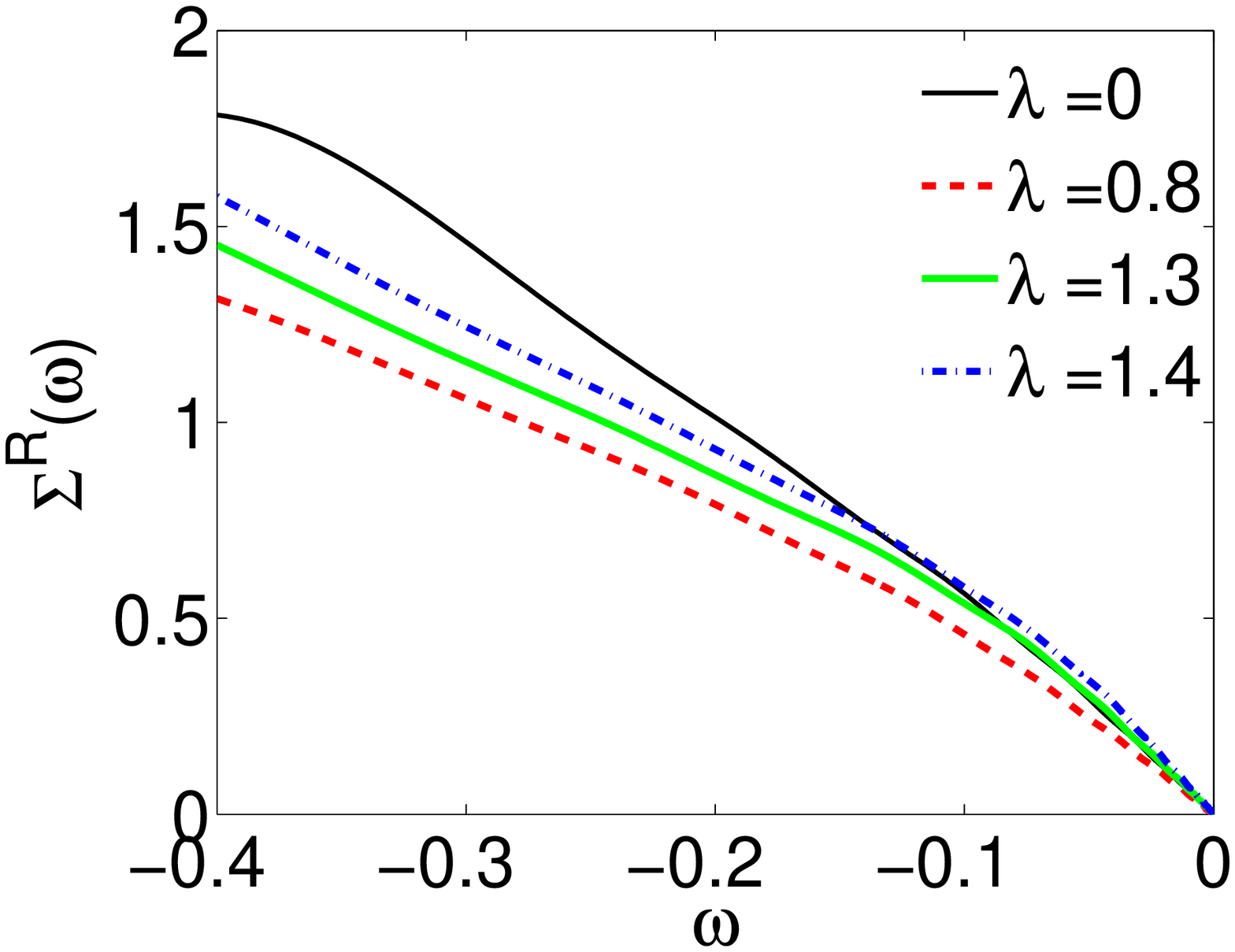}}
\subfigure[]{\includegraphics[width=0.23\textwidth]{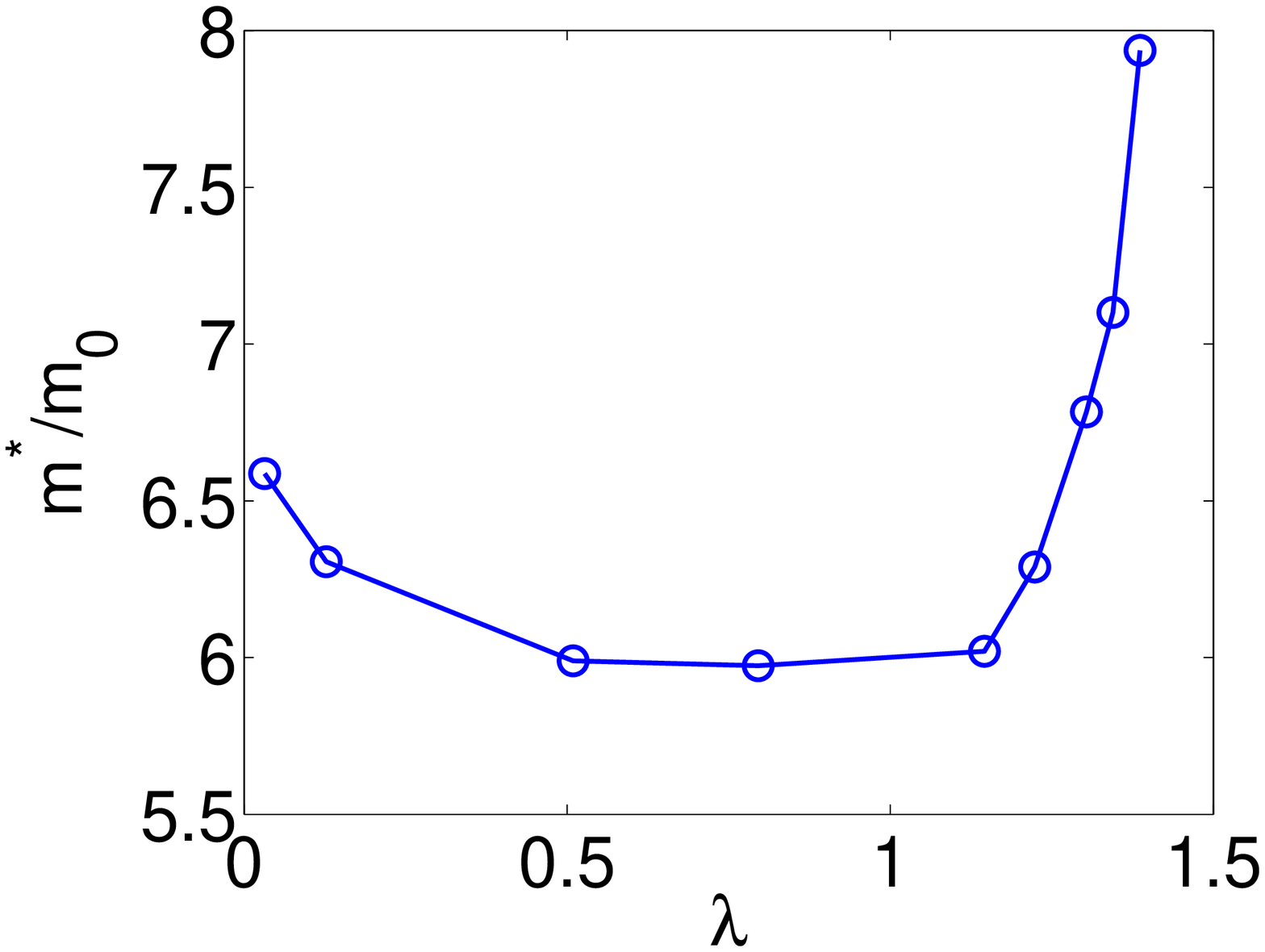}}
\caption{(Color online) PM calculation: (a) The $\epsilon_{\vk}$ resolved
  spectral function for $U=6$, $n=0.9$ and $\lambda=0.86$. Note that for better
  visibility of all the peaks, the largest ones have been cut off at a certain
  value. This generates a small artificial plateau, otherwise absent. (b) The
  dispersion relation $E_{\vk}$ as a function of 
  $\epsilon_{\vk}$  for $U=6$ and $\lambda=0$ and $\lambda=0.86$. The
  horizontal dotted (dashed) line gives the energy $\omega_0$ ($\omega_0^r$),
  and the dashed line is for free electrons, $E_{\vk}=\epsilon_{\vk}$. 
 (c) The real part of
  the self-energy $\Sigma^{\rm  R}(\omega)$ (with $\Sigma^{\rm  R}(0)$
  subtracted) as a function of $\omega$  for $U=6$ and various values of
  $\lambda$. (d) The effective mass as a function of $\lambda$.}        
\label{resultsU6}
\end{figure}

\noindent
One can observe in Fig. \ref{specU6} that the central quasiparticle peak
is relatively little affected by the presence of the electron-phonon coupling.  
In fact, the effect of the electron-phonon interaction on the low-energy
quasiparticle appear just as a small ``de-renormalization''; in other words
a \emph{decrease} in the effective mass $m^*$ (see Fig. \ref{resultsU6}d for
small values of $\lambda$). 
This result has been already observed at half-filling
within DMFT using Lanczos as impurity solver \cite{StJ2}. It can be understood
from the physics of the effective impurity model and the Kondo effect
connected with the quasiparticle peak. It comes from the fact
that the electron-phonon interaction effectively reduces the bare repulsion of
the system, making the Kondo coupling somewhat larger.
Such a physical mechanism can also be described within perturbation theory,
assuming that the scale $U$ associated to the separation between the Kondo
peak and the Hubbard bands is the largest, compared to $t$ and $\omega_0$
\cite{grempel}. 
It is remarkable that we observe this de-renormalization effect so clearly
even away from half-filling, i.e. where the abovementioned separation of
energy scales is no longer clearcut (at least for the removal part of the
spectrum). 
Moreover this result convincingly shows how this kind of ``protection'' of the
low-frequency part of the spectrum from phonon effects characteristic of the
paramagnetic DMFT solution is present also within the NRG solution.
At higher energies instead, one can see differences between the
solution with and without $\lambda$. Here, contrary to the low-frequency regime,
the resolution of our NRG calculation is not accurate enough to resolve the
details of the spectra, because of the conventional high-energy
broadening used.

Let us now turn to the discussion of the dispersion kink.
In Fig. \ref{resultsU6} we show a collection of results for the situation with
the same $U$ and $n$ and different values of $\lambda$. 
The low energy part of $\rho_{\vk}(\omega)$ for
$\lambda\!=\!0.86$ is shown in Fig. \ref{resultsU6} (a) and the corresponding
dispersion relation in panel (b).

\noindent
Comparing the result for $\lambda\!=\!0$ and $\lambda\!=\!0.86$, we can see how
the whole behavior is modified due to the electron-phonon coupling even though
no clear kink is visible. 
We just get a continuous change of the slope for relatively large values of the coupling.
We can understand this in light of the arguments given above:
the Holstein electron-phonon interaction can not easily induce charge
fluctuations due to the direct competition with the Hubbard $U$. 
Even though the finite doping reduces the influence of the Mott physics
compared to half-filling \cite{KMH04,SCCG05}, the whole low-energy (Kondo-like)
physics is still dominated by the electron-electron interaction and room for
nontrivial electron-phonon effects is left only for higher-frequency
(atomic-like) processes.  
This has two very important consequences: (i) the low-frequency slope of the
dispersion is not strongly affected by the bare electron-phonon coupling and
(ii) the deviation between the dispersion with and without phonons (either in
the form of a sharp kink or as a smooth change of slope) becomes a
\emph{high-frequency} \emph{electronic} property, in striking contrast with
the standard uncorrelated electron-phonon picture in which this is bound to
the (renormalized) phonon frequency.

The $\omega$-dependence of the real part of the self-energy, shown in panel
(c), consequently shows relatively little modification at low energy and visible one
at higher energy.  
The behavior of the low energy slope and correspondingly effective mass is
shown in part (d) as function of $\lambda$.  As we noted above, there is an initial
de-renormalization tendency upon increasing $\lambda$, followed by a 
strong increase 
of $m^*$ for larger values of $\lambda$. 

\begin{figure}[!thbp]
\centering
\includegraphics[width=0.45\textwidth]{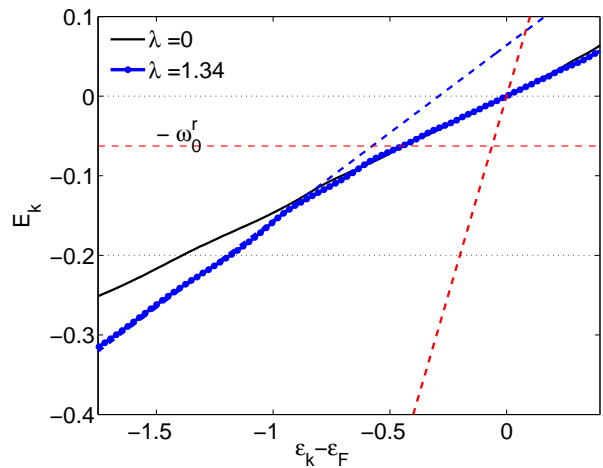}
\caption{(Color online) PM calculation: The dispersion relation $E_{\vk}$ as a function of $\epsilon_{\vk}$
  for $U=6$, $n=0.9$ and different values of electron-phonon coupling. Included is a
  dashed line to fit the high energy slope of the dispersion.}       
\label{dispU6g0.65}
\end{figure}

We noted for Fig. \ref{resultsU6} (b) that no kink feature can easily be
identified. This changes when we increase $\lambda$.
In Fig. \ref{dispU6g0.65} we show the dispersion at
a larger value $\lambda$=$1.34$.  

\noindent
A kink feature is visible at $\omega_k$$\approx$$0.15$ while the renormalized phonon
frequency $\omega_0^r$$\approx$$0.063$ (extracted from the maximum of the phonon spectral
function $\rho_b(\omega)$ see inset Fig. \ref{phonsoftU0}).  
This clearly shows how these two quantities, always bound to go hand in hand
in a conventional uncorrelated metal as two aspects of the same physical
phenomenon, can instead be different in a system which is described within
paramagnetic DMFT.  
The renormalized phonon frequency $\omega_0^r$ still describes the influence
of electron-phonon coupling on the phonon spectral function, while the kink
position $\omega_k$ is associated to the energy range in the electronic
spectrum where phonon effects become visible.  
The latter is a new energy scale of the problem, which comes out of the
competition between the instantaneous Hubbard repulsion and the retarded
phonon-mediated attraction. It is interesting to see that, as seen  in the
inset of Fig. \ref{phonsoftU0}, $\rho_b(\omega)$, which for
$\lambda=0$ is a delta function peaked at $\omega_0$, possesses a 
considerable amount of spectral weight for energies larger than $\omega_0^r$
for this situation. By numerical fitting the magnitude of the kink was
extracted to be $m_k\approx 1.55$. When we increase $\lambda$ further to
overcome $U$, strong polaronic behavior will eventually dominate the low
energy behavior again.  

\begin{figure}[!thbp]
\centering
\includegraphics[width=0.45\textwidth]{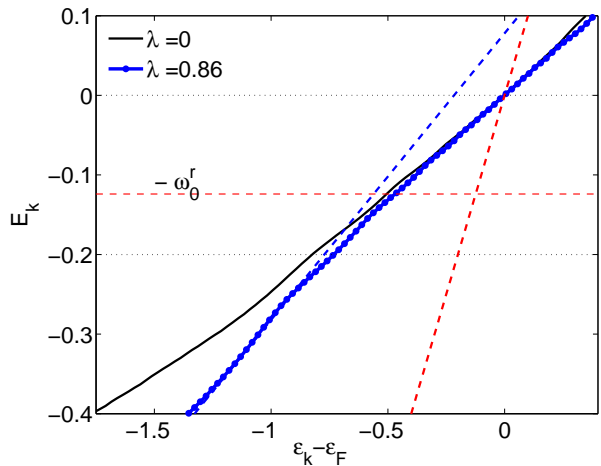}
\caption{(Color online) PM calculation: The dispersion relation $E_{\vk}$ as a function of $\epsilon_{\vk}$
  for $U=6$, $n=0.8$ and different values of electron-phonon coupling. Included is a
  dashed line to fit the high energy slope of the dispersion.}       
\label{dispU6g0.5x0.8}
\end{figure}

Another possibility to find a kink due to electron-phonon coupling in the
presence of large $U$ is to increase the doping, as the effect of $U$ to
freeze the charge fluctuations is weakened then. We give an example for $\lambda=0.86$ and 20$\%$
doping in Fig. \ref{dispU6g0.5x0.8}. 

\noindent
A weak kink feature can be identified at higher energy. This is more
pronounced in calculations for larger values of $\lambda$, but we stick to this value for
comparison to later results. The magnitude of the kink is by numerical
fitting $m_k\approx 1.39$.
Notice that also here the kink position is  larger than the
renormalized phonon energy. When we further increase the doping $m_k$ will
become larger and $\omega_k$  will, for $n\sim 0.5$, eventually coincide\cite{KHE05} with 
$\omega_0^r$ as expected in the conventional picture.

We cannot compare the DMFT results obtained so far to experiments on cuprates
as we have not yet introduced any effect coming from antiferromagnetic
correlations.  
These are known to play a crucial role for small doping values, therefore we
have to include them in our model calculation. 
One way of doing this is to let the system have long-range order so that we can
describe (a classical mean-field version of) the physics of antiferromagnetic
exchange. This is what is done in the following section. 

\section{Kinks and antiferromagnetic correlations}
\label{sec:kinkAFM}
In order to include antiferromagnetic correlations we assume a bipartite
lattice with $A$ and $B$ sublattice and, as in the standard DMFT
implementation for long-range ordered phases
\cite{GKKR96,ZPB02,AFDMFT,BH07c}, we write the matrix Green's function in
the form      
\begin{equation}
\underline{G}_{\vk,\sigma}(\omega) \!=\!
\frac1{\zeta_{A,\sigma}(\omega)\zeta_{B,\sigma}(\omega) -\epsilon_{\vk}^2}
\! \left(\!\!\!
\begin{array} {cc}
 \zeta_{B,\sigma}(\omega) & \epsilon_{\vk} \\
\epsilon_{\vk} & \zeta_{A,\sigma}(\omega)
\end{array}
\!\!\!\right),
\label{afmkgf}
\end{equation}
with $\zeta_{\alpha,\sigma}(\omega)=\omega+\mu_{\alpha,\sigma}-\Sigma_{\alpha,\sigma}(\omega)$,
$\alpha=A,B$.
For the
AF order one has $\mu_{A,\sigma}=\mu-\sigma h_s$, $\mu_{B,\sigma}=\mu+\sigma
h_s$, and the condition
$\Sigma_{B,\sigma}(\omega)=\Sigma_{A,-\sigma}(\omega)$.   
We consider solutions where the symmetry breaking field vanishes, $h_s\to 0$.
As we do not have any next-nearest neighbor hopping, the antiferromagnetic
solution is lower in energy than the paramagnetic one for small values of
the doping \cite{ZPB02}. 

The form of the dispersion as found from the poles of the Green's function in 
Eq. (\ref{afmkgf}) can be obtained from the equation
\begin{equation}
  \omega_{\pm}=-\bar\mu+\Sigma^+(\omega)\pm\sqrt{(\Delta\bar\mu-\Sigma^-(\omega))^2+\epsilon_{\vk}^2},
\label{afmdisp}
\end{equation}
where  we only included the real parts of the self-energies.
We defined $\bar \mu=(\bar\mu_{\uparrow}+\bar\mu_{\downarrow})/2$, $\Delta\bar
\mu=(\bar\mu_{\uparrow}-\bar\mu_{\downarrow})/2$ with
$\bar\mu_{\sigma}=\mu_{\sigma}-\Sigma_{\sigma,0}$, and we have split into
static and dynamic part in the self-energy,
$\Real\Sigma_{\sigma}(\omega)=\Sigma_{\sigma,0}+\bar\Sigma_{\sigma}(\omega)$. 
We also introduced
$\Sigma^{\pm}(\omega)=(\bar\Sigma_{\uparrow}(\omega)\pm\bar\Sigma_{\downarrow}(\omega))/2$. 
The solution of Eq. (\ref{afmdisp}) as an implicit equation gives the dispersion
$E_{\vk,\pm}$ in the AF case. The mean field solution without phonons gives 
the two Slater bands, and upon hole doping the Fermi level moves into the
lower one. By making a simple ansatz for the electron-phonon part of the
self-energy,
$\bar\Sigma_{\sigma}^g(\omega)=-\lambda_{\sigma}\omega\theta(\omega_0-|\omega|)$
with $\lambda=(\lambda_{\uparrow}+\lambda_{\downarrow})/2$,
one can show that the different form of the equation does not lead 
to the occurrence of larger kinks in the dispersion relation than in the PM
case, Eq. (\ref{dispeq}).  

Several studies pointed out the importance of cooperative effects between
lattice- and spin-polarons \cite{volly,AFDMFT}, 
influencing both photoemission
\cite{fehske-tJ,ciuchi-tJ,mishchenko,olle2,cataudellaPRL99,olleBorn,karPRB78}  
and optical properties \cite{fehskeOpt,Mea08,cappellutiPRB79} of correlated systems.
By studying antiferromagnetism in the Hubbard-Holstein model within DMFT it
was clarified that a finite doping weakens 
these cooperative effects \cite{StJ3}.
However, the values of $\lambda$ at which polarons are obtained in the
antiferromagnetic phase are not only much smaller than 
the values needed to form local bipolarons but are also smaller than those needed to observe polaronic
features in paramagnetic DMFT spectra \cite{StJ3}.
It is interesting to note that Dynamical Cluster Approximation (DCA)
calculations \cite{macridin} confirmed the picture found in AF-DMFT, even
though short-range AF correlations in DCA are realistically present already at
the paramagnetic level. 

\begin{figure}[!thbp]
\centering
\includegraphics[width=0.45\textwidth]{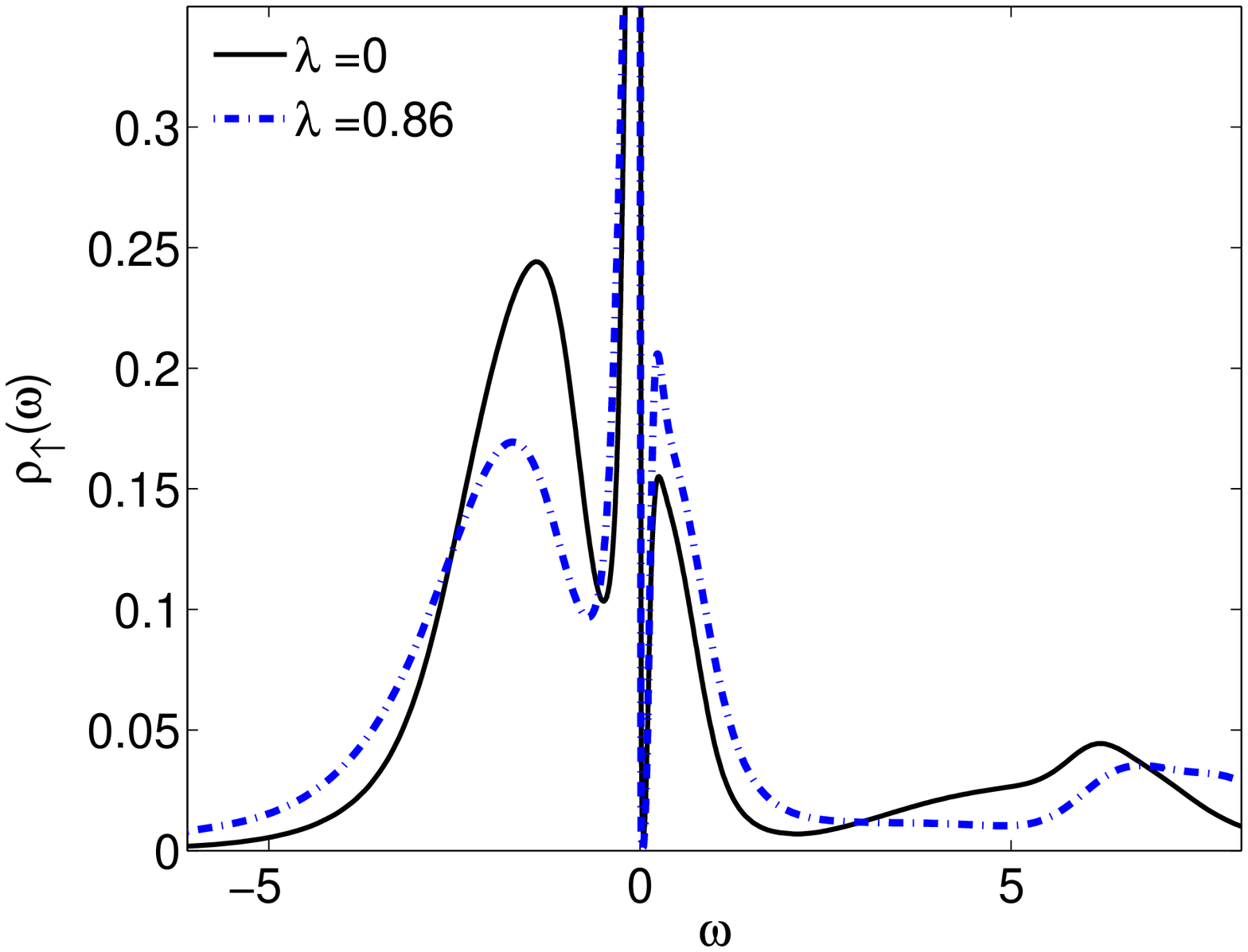}
\includegraphics[width=0.45\textwidth]{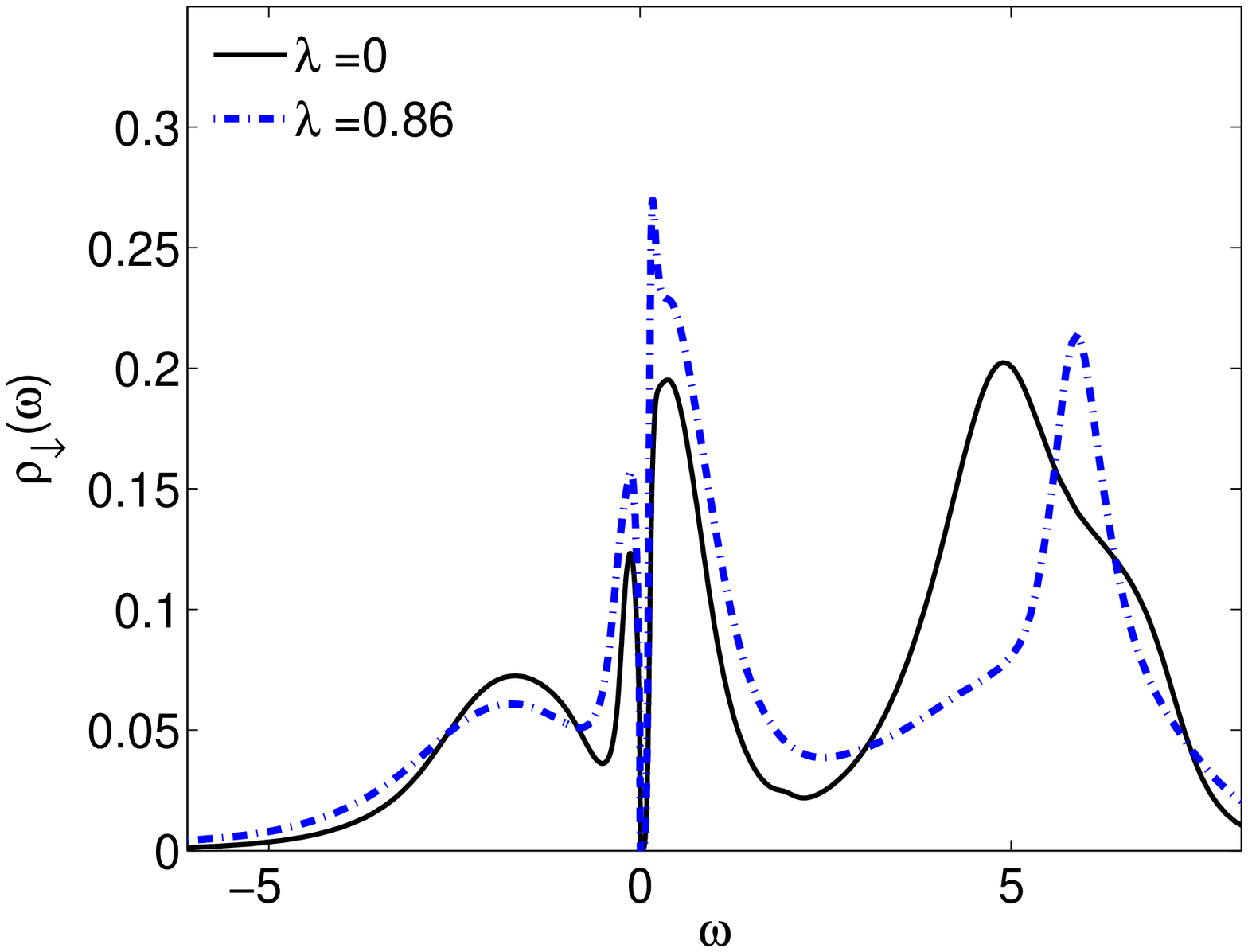}
\caption{(Color online) AF calculation: The local electronic spectral density for majority electrons
  $\rho_{\uparrow}(\omega)$ and minority electrons $\rho_{\downarrow}(\omega)$.}
\label{specafmU6}
\end{figure}

The present AF-DMFT NRG study has the big advantage that we can
resolve low-energy features in the spectral function at finite doping in a way
that is beyond any of the previous exact diagonalization and Monte Carlo
studies. 
In Fig. \ref{specafmU6} we show the spin-resolved local spectral function in
AF-DMFT solved with NRG.  
The parameters, $U$=$6$ and $n$=$0.9$, are the same as in
Fig. \ref{specU6} and \ref{resultsU6}(b) where the spectral function without
antiferromagnetic correlations was shown. 
\noindent
As one can see the Fermi energy has moved into the lower band and at low
energy the typical square root divergence in $\rho_{\uparrow}(\omega)$
appears. At the same time the minority spin $\rho_{\downarrow}(\omega)$
displays the square root decrease. Hubbard band features at higher energies can also be identified. 

\begin{figure}[!thbp]
\centering
\includegraphics[width=0.45\textwidth]{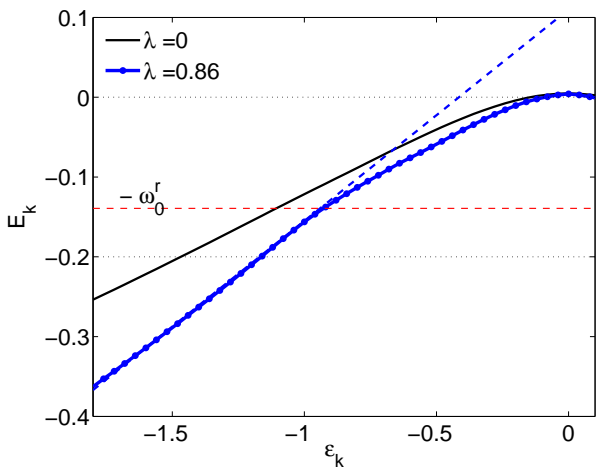}
\caption{(Color online)  AF calculation: The dispersion relation $E_{\vk}$ as a function of $\epsilon_{\vk}$
  for $U=6$, $n=0.9$ and different values of electron-phonon coupling. Included is a
  dashed line to fit the high energy slope of the dispersion.}       
\label{dispafmU6g0.5}
\end{figure}

The sublattice spectral functions of the AF solution at 10$\%$ doping and $U=6$ 
show qualitatively similar features for $\lambda=0$ and $\lambda=0.86$ as seen in
Fig. \ref{specafmU6}.
The spectrum at $\lambda$=$0.86$ displays a strong quasiparticle
excitation at the gap edge, with no big sign of 
renormalization.  
Yet the effect of the electron-phonon interaction is still much stronger than
in the paramagnetic DMFT at the same $U$.  
To see this we show in Fig. \ref{dispafmU6g0.5} the dispersion relation for
the same set of parameters, comparing again $\lambda$=$0$ with
$\lambda$=$0.86$. 

\noindent
A kink in $E_{\vk}$ is visible, while (see
Fig. \ref{resultsU6}b) just a smooth change in the slope could be observed for
the same value of $\lambda$ without the AF correlations. By numerical fitting
we can extract a magnitude of the kink of $m_k$$\approx$$1.64$. When doping is
decreased or $\lambda$ increased the kink becomes even more pronounced. Similar values of
$m_k$ within the paramagnetic phase are reached only taking
$\lambda$=$1.34$. Therefore, the 
qualitative difference between paramagnetic and AF-DMFT coming from the
underlying spin background reflects in a big quantitative difference in the
strength of the kink. This is expected from the above arguments, but has not
been shown explicitely for the dispersion relation.

Particularly interesting for the present study is the kink position $\omega_k$ in AF-DMFT.
Contrary to the paramagnetic phase studied in Sec. \ref{sec:PM}, the kink in
the electronic dispersion is \emph{at} (or even slightly before) the
renormalised frequency $\omega_0^r\simeq 0.16$, as shown in
Fig. \ref{dispafmU6g0.5}.  This is very different from the paramagnetic DMFT
result of the previous section, for which $\omega_k$ looses its
connection with the renormalized phonon frequency and becomes a higher
frequency electronic property.  
For the values of $\lambda$ or doping range where a clear kink can be identified
($\lambda$=$1.34$, $n\le0.8$), indeed, paramagnetic DMFT gives a renormalized phonon
frequency $\omega_0^r<\omega_k$.
Therefore we conclude that paramagnetic and
antiferromagnetic DMFT calculations can differ qualitatively in the physics behind
the kink: in paramagnetic DMFT a larger $\lambda$ or larger doping is needed to
form a kink and the kink energy is not pinned to the renormalized phonon frequency, but rather
is connected to a higher energy scale at the crossover between the Kondo-like
physics and a more atomic-like one. In AF-DMFT a smaller value of $\lambda$ is
enough to see a kink and the position of this kink closely tracks the
renormalized phonon frequency.

\begin{figure}[!thbp]
\centering
\includegraphics[width=0.45\textwidth]{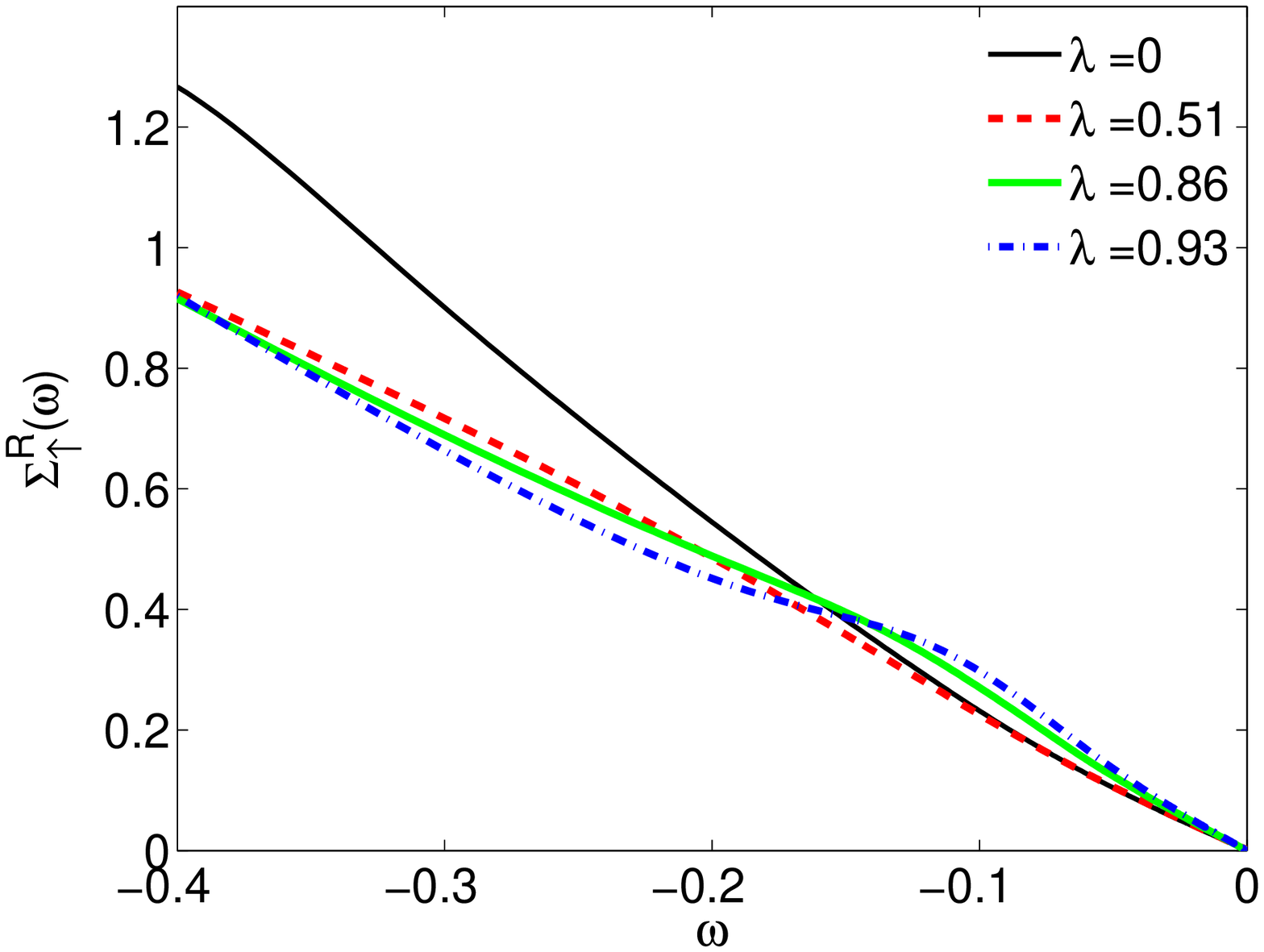}
\includegraphics[width=0.45\textwidth]{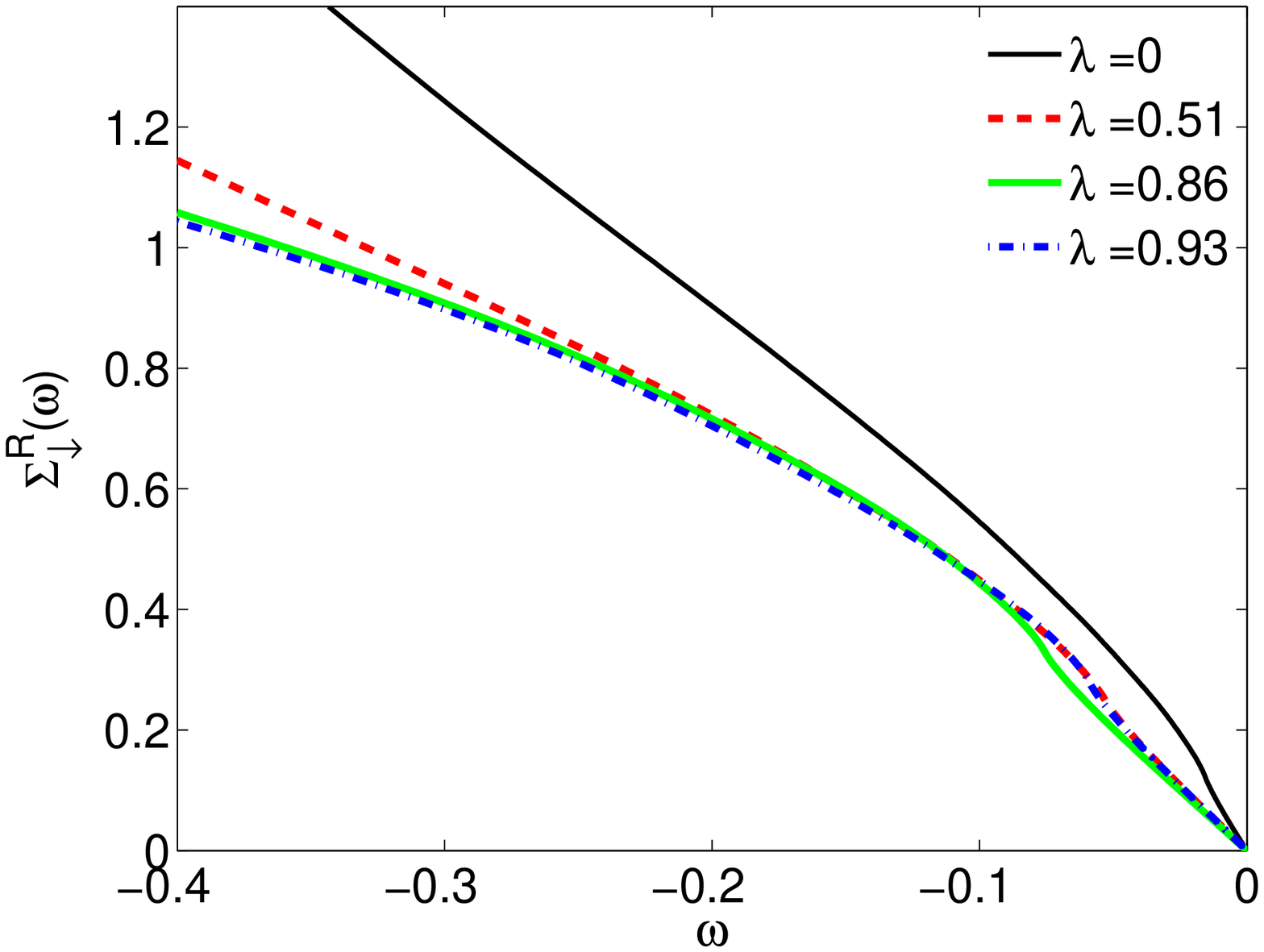}
\caption{(Color online)  AF calculation: The real part of the selfenergies $\Sigma^{\rm  R}_{\uparrow}(\omega)$ (upper panel) and
  $\Sigma^{\rm  R}_{\downarrow}(\omega)$ (lower panel) with $\Sigma^{\rm  R}_{\sigma}(0)$ subtracted.}       
\label{afmselfenU6}
\end{figure}

The stronger electron-phonon effects on the AF electronic dispersion are
connected directly to the real part of the self-energy, shown in
Fig. \ref{afmselfenU6} for the majority ($\uparrow$) and minority ($\downarrow$) spin component on the sublattice.
 
\noindent
Without the electron-phonon coupling minority electrons
scatter much stronger with the majority spin electrons and therefore the slope
of $\Sigma^R_{\downarrow}(\omega)$ is larger than the one of
$\Sigma^R_{\uparrow}(\omega)$ for $\lambda=0$.\cite{BH07c}
When $\lambda$ is increased, we can see that 
the low energy slope changes relatively little for $\Sigma^R_{\uparrow}(\omega)$.  
At $\lambda\ge 0.86$ a feature at $\omega\simeq\omega_0^r$ can be identified
in $\Sigma^R_{\uparrow}(\omega)$, which gives rise to the kink in the
dispersion in Fig. \ref{dispafmU6g0.5}. In calculations for larger $\lambda$, smaller doping
or smaller $U$ this is more pronounced. 
This is related to the cooperative effect of the magnetic polaron
interacting with phonons which was discussed
earlier. The low energy slope of $\Sigma^R_{\downarrow}(\omega)$ decreases on
increasing $\lambda$ and no clear feature is seen around $\omega\simeq\omega_0^r$ .

For small values of $U$ and $\lambda$ these results for the self-energies can
be understood from a perturbative calculation to order $g^2$ (see
Fig. \ref{g2diagram}). As a starting point one can use the  doped
AF state and calculate the lowest order self-energy corrections
due to coupling to the phonons. Then it is found that the majority spin electrons
scatter much stronger with the phonons as the minority ones, and are therefore
much stronger renormalized. $\Sigma^R_{\uparrow}(\omega)$ shows a much more pronounced logarithmic
divergence at $\omega^r_0$ than in the normal state discussed in
Sec. \ref{sec:kinkPM}.  Formally, this  
comes from the enhanced density of states at the Fermi level which is a
consequence of the AF order. An intuitive understanding can be motivated by
considering that majority spin electrons on a sublattice are surrounded on
average by the opposite spin direction.  This provides an energetic gain, but
also leads to some degree of localization, as an itinerant excitation destroys
the antiferromagnetic order.  This effect can cooperate with the coupling to the phonons
inducing a further renormalization and as a result majority spins are more
affected than without the ordered background.  At the same time minority spin
electrons couple less to the phonons. In addition to the different renormalization effects, the
electron-phonon coupling can lead to a stabilisation and even an increase of
the magnetization. These weak coupling considerations qualitatively carry over
to the stronger coupling case relevant for the results above. However,
additional renormalization effects due to $U$ play a role then and make a
perturbative treatment difficult.

\section{Conclusions }
\label{sec:Concl}
We have described the occurence of kinks in the electronic dispersion relation
due to coupling to a local phonon mode taking into account strong electronic
correlations. As anticipated our results show that the simple theory of kinks
is modified for both strong electron-phonon and electron-electron
coupling. For the paramagnetic state and small doping from half filling a
clear kink is not visible anymore for large values of $U$$>$$W$ unless the electron-phonon 
coupling assumes large values. The phonons can have an effect at higher energy
whilst the low energy features are less affected. An important finding is that
the effect of $U$ to suppress the occurrence of kinks is less
efficient if we allow for antiferromagnetic correlations. Then a cooperative
effect leads to a stronger effective coupling of the electronic quasiparticles
(magnetic polarons) to the phonons. That something like this can happen has
been observed in earlier studies \cite{StJ3,MMJM09}. Here we clarify how this
manifests itself for kinks. Nevertheless for the type of phonons we study in
competition with the Coulomb interaction we find that inspite of relatively
large values, $\lambda\simeq 0.8$, kink features are moderate even with
antiferromagnetic correlations, $m_k\simeq 1.6$.
This may be a consequence of the type of phonons we study (Holstein).
Some works have indeed shown that non-Holstein phonons have stronger polaronic
effects in correlated systems, due to a less direct competition with the local Hubbard 
repulsion \cite{olle1,defilippisKink,johnstonPRB82}. 
Therefore a quantitative comparison with results for the low-energy kink in
cuprate superconductors, or a statement about the origin of the nodal kink in
these systems is hard to make here. 

Still, we can relate our findings to the puzzling results for the kink
position reported by Kordyuk {\it et al.} in Ref. \onlinecite{kordyukPRL97}.
There, as already discussed in the Introduction, a suprisingly strong
dependence on doping and temperature of the kink position was observed.  
Nothing like this is can be obtained within conventional electron-phonon theories. 
Our results show that, by taking the effect of electronic correlation
properly into account, the kink position becomes strongly dependent on the
strength of antiferromagnetic correlations.  We found that when AF
correlations are strong, the kink is close to the renormalized phonon
frequency (i.e. is more ``low energy''), while when AF correlations are
absent, the kink can occur at higher frequencies and also needs a larger
electron-phonon coupling to be of similar strength. 
Therefore it is very plausible to expect that, at low doping, where AF
correlations are stronger, the kink energy is small in absolute value (see
Fig. \ref{dispafmU6g0.5}), while
at larger values of the doping, where AF correlations are weaker, the kink
occurs at larger energies (see Fig. \ref{dispU6g0.5x0.8}). Interestingly, this is
exactly the trend shown in the data on the Bi-compound at the lowest
temperature \cite{kordyukPRL97}.  

Generally, it is very important to test - as we do here - the predictions of theories
which, like paramagnetic and AF-DMFT, describe electron-phonon \emph{and}
strong electronic correlations on an equal footing.  
Strong coupling to a phonon mode is very often discarded as being 
responsible for the nodal kink in cuprates on the basis of clich{\'e}s coming
from conventional theories neglecting electronic correlations. 
The doping dependence is one typical example.
Our study shows instead that, when including correlation, properties
like the doping dependence of the kink energy may become highly non-trivial,
and therefore the conventional picture cannot be used to rule out phonons.

\bigskip
\noindent{\bf Acknowledgment}\par
\noindent
We wish to thank O. Gunnarsson, A.C. Hewson and D. Manske for useful discussions and
O. Gunnarsson for critically reading of the manuscript. G.S. acknowledges financial support from the FWF
'Lise-Meitner' Grant No. M1136.


\bibliography{artikel,biblio1,added,newBIB}

\end{document}